\documentclass[aps,pra,twocolumn,showpacs,preprintnumbers,amsmath,amssymb,superscriptaddress,10pt,nofootinbib]{revtex4-1}

\usepackage{amsmath,amssymb}
\usepackage{bm}
\usepackage[latin1]{inputenc} 
\usepackage{dcolumn}
\usepackage{graphicx}
\usepackage{SIunits}
\usepackage{multirow}
\usepackage{ae}
\usepackage{subfigure}
\usepackage{hyperref}
\usepackage{color} 

\newcommand{\matrise}[1]{\begin{bmatrix} #1 \end{bmatrix}}
\newcommand{\diff}[0]{\text{d}}

\renewcommand{\Im}{\mathrm{Im}}

\newcommand{\vek}[1]{\boldsymbol{\mathbf{#1}}}
\newcommand{\vekh}[1]{\hat{{\boldsymbol{\mathbf{#1}}}}}
\renewcommand{\div}[0]{\nabla\cdot\vek}

\newcommand{\curl}{\nabla\times\vek}

\newcommand{\be}{\begin{equation}}
\newcommand{\ee}{\end{equation}}

\begin{document}

\title{Reciprocity and the scattering matrix of waveguide modes}

\author{Guro K. Svendsen}
\affiliation{Department of Electronics and Telecommunications, Norwegian University of Science and Technology, NO-7491 Trondheim, Norway}

\author{Magnus W. Haakestad}
\affiliation{Norwegian Defence Research Establishment (FFI),
NO-2027 Kjeller, Norway}

\author{Johannes Skaar}
\affiliation{Department of Electronics and Telecommunications, Norwegian University of Science and Technology, NO-7491 Trondheim, Norway}
\email{johannes.skaar@ntnu.no}

\date{\today}

\begin{abstract}
The implications of the Lorentz reciprocity theorem for a scatterer connected to waveguides with arbitrary modes, including degenerate, evanescent, and complex modes, are discussed. In general it turns out that a matrix $CS$ is symmetric, where $C$ is the matrix of generalized orthogonality coefficients, and $S$ is the scattering matrix. Examples are given, including a scatterer surrounded by waveguides or free space, and discontinuities of waveguides.
\end{abstract}

\maketitle

\section{Introduction}
The Lorentz reciprocity theorem is a famous and very useful result in electromagnetics \cite{lorentz, snyderlove, pozar, vassallo1977, carminati1998}. In particular, it has important consequences in scattering theory, relating the transmission coefficients when the position of the source and detector are interchanged. For orthogonal, normalized, propagating modes in waveguides or free space, the forward and backward transmission coefficients due to a scatterer with symmetric permittivity and permeability tensors, are equal. This symmetry has proved to be useful in a number of situations in physics and electrical engineering. For example, the symmetry implies that the transmission coefficients through windows, thin-film filters, or fiber Bragg gratings are equal from opposite sides. The symmetry property also shows that one-way components such as isolators and circulators, cannot be made of reciprocal materials.

With the demand for device miniaturization, coupling to and from evanescent modes becomes increasingly important. For instance, when considering the end-facet reflection and transmission of nanowire lasers, or the reflection and transmission of nanowire Bragg reflectors, the evanescent waveguide modes play an important role \cite{svendsen2011,svendsen2012}. For certain metal-dielectric waveguides, there are modes with complex propagation constants \cite{clarricoats64,sturman07}, even though the material is lossless. In the presence of degeneracy, modes may not be orthogonal. With the recent interest in plasmonic, metamaterial, and photonic crystal waveguides it is thus of interest to see if there are symmetry relations for the scattering matrix when complex and/or degenerate waveguide modes are included. 

While symmetry relations are well known in the case of propagating modes, the case with evanescent modes is less developed in the literature. Vassallo discussed the symmetry relation for the scattering matrix in the presence of evanescent waveguide modes \cite{vassallo1977}. For the special case with free space on each side of the scatterer, Carminati et al. found the reciprocity relation for the scattering matrix associated with plane, electromagnetic waves \cite{carminati1998}. Earlier treatments did not include the possibility of complex modes, and nonorthogonal modes.

Here we will provide a self-contained treatment of the Lorentz reciprocity theorem and its implication for the scattering matrix or transmission coefficients between propagating, evanescent, and complex modes. We also consider the effect of degeneracy in detail. The scatterer is contained in a black box with one or more waveguide inputs, and/or free space on one or both sides. As stated in detail in the next section, the analysis is quite general; it only requires the usual assumptions for the Lorentz' reciprocity theorem to be valid for the scatterer, and the additional assumption that the input/output waveguides are described by scalar permittivities and permeabilities. Examples are given, which demonstrate the reciprocity relations for a discontinuity between two waveguides.

It will turn out that it is not the scattering matrix $S$, but the product $CS$, that is symmetric. Here $C$ is the matrix of generalized orthogonality coefficients \eqref{orthogonality}. This result is useful when designing or analyzing structures with nontrivial waveguides, including plasmonic or nanowire waveguides. For example, when designing waveguides with Bragg gratings or other inhomogeneities along the waveguide axis, it is useful to relate the reflection and transmission coefficients from the two sides, in order to identify transfer matrices of the different sections. Also, it is important to have the symmetry in mind, to be aware of the inherent limitations for structures satisfying the conditions of the Lorentz' reciprocity theorem.

\section{Lorentz reciprocity and the scattering matrix}
Consider a volume $v$ with a linear, time-invariant, and spatially nondispersive medium. The medium may be lossy, temporally dispersive, anisotropic, and inhomogeneous. Let $\epsilon$ and $\mu$ be the permittivity and permeability tensors in $v$, respectively. We assume that $\epsilon$ and $\mu$ are symmetric; this property follows from thermodynamic relations under very general conditions \cite{landau_lifshitz_edcm}. The conductivity, if any, is included in the (possibly complex) permittivity $\epsilon$. The volume $v$ is seen as a black box with an unknown system; however in addition to the assumptions above it is known that there are no current sources in $v$. Let \{$\vek E^1$, $\vek H^1$\} and \{$\vek E^2$, $\vek H^2$\} be the electromagnetic fields in $v$, resulting from excitations 1 and 2, respectively, whose sources are located outside $v$. The Lorentz reciprocity theorem reads (see Appendix A):
\begin{equation}\label{lorentzrec}
 \oint_s (\vek E^1\times\vek H^2)\cdot\diff\vek S=\oint_s (\vek E^2\times\vek H^1)\cdot\diff\vek S,
\end{equation}
where the closed surface $s$ is the boundary of $v$.

We now consider a situation where the volume $v$ has several input/output waveguides (Fig. 1).
\begin{figure}
\includegraphics[width=7cm]{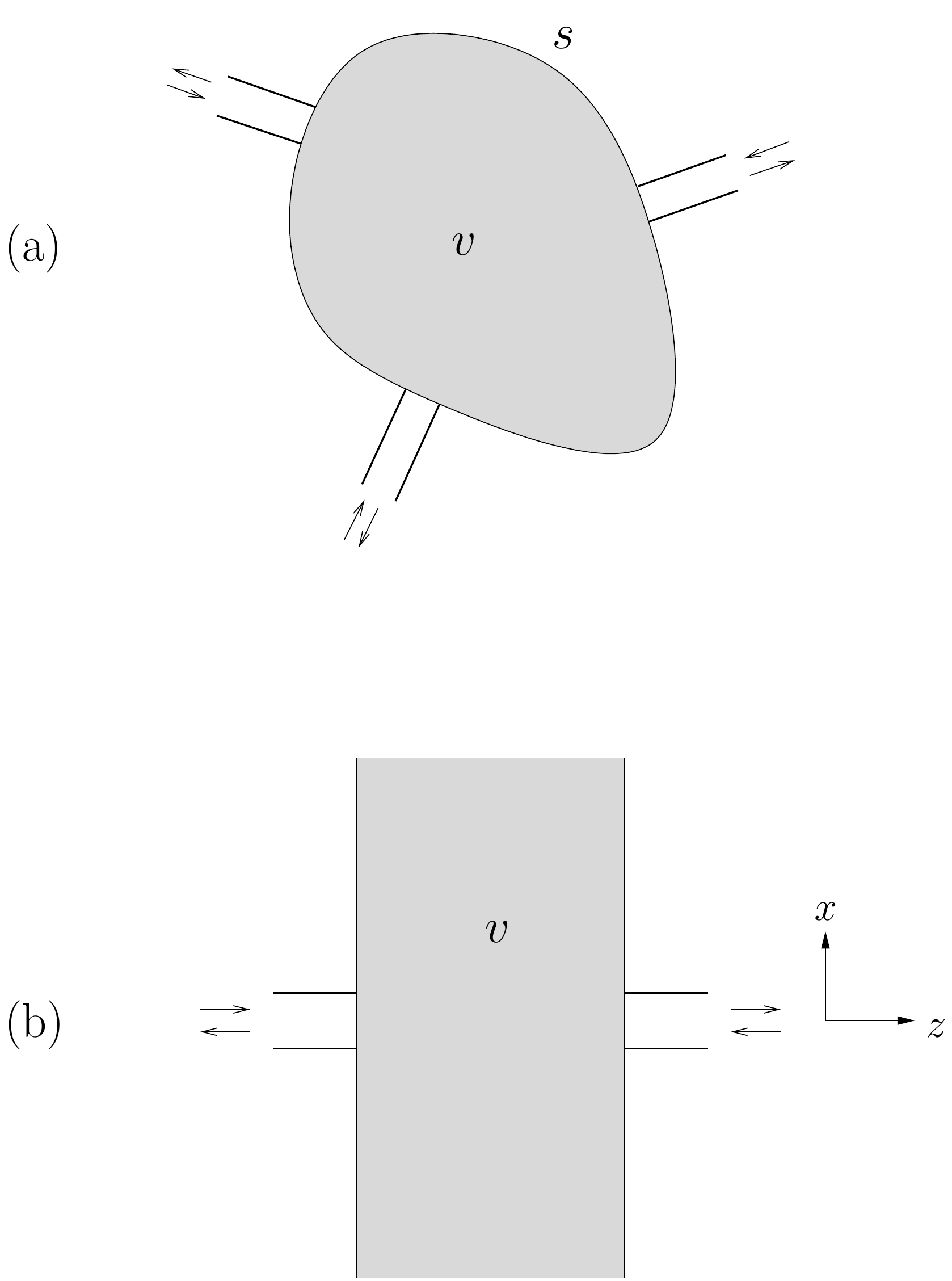}
\caption{The ``black box'', or scatterer, is contained in a volume $v$ with boundary $s$. (a) The general case with several waveguides connected to the surface $s$. (b) The special case with two waveguides, one on each side of $v$. The surface $s$ is regarded as two infinite planes.} 
\label{fig:reciprocity}
\end{figure}
Each of the waveguides is described by a permittivity and permeability which are uniform along the waveguide's axis. The waveguides may be lossless or lossy. We assume that any solution to Maxwell's equations in the waveguides are expressible as a superposition of modes: Bound modes, radiation modes, and/or evanescent and complex modes. The waveguides are assumed to be isolated, in the sense that their modal fields do not overlap.

An excitation in one or more waveguides may lead to forward and backward propagating waves in each mode. Here, ``forward'' refers to the input direction, while ``backward'' refers to the output direction. A mode is referred to as ``forward propagating'' if its power flow is directed in the forward direction (for propagating modes) or if it decays in the ``forward'' direction (evanescent or complex modes), see Appendix \ref{waveguidemodes}. Let the modal amplitudes of the forward and backward propagating waves be $a_j$ and $b_j$, respectively. This set includes all modes in all waveguides. The modes are labeled by $j$, which is discrete in the discrete part of the mode spectra, and continuous otherwise. Alternatively, we can ensure that the entire mode spectra are discrete by imposing artificial boundary conditions, such as periodic or metallic boundary conditions. For example, if a dielectric waveguide is surrounded by a metallic box, the continuous set of radiation modes gets discretized.

When two or more waveguides do not have parallel axes (Fig. 1(a)), there is a fundamental problem with the radiation modes associated with dielectric waveguides. In principle, such modes extend infinitely in the transversal waveguide direction, and will therefore overlap spatially, as opposed to our earlier assumption about isolated waveguides. 
There are at least two remedies. The first remedy is to consider supermodes rather than modes of separate waveguides. In other words, one considers the true modes of the composite structure consisting of two or more waveguides. The second, perhaps more practical remedy is again to use artificial boundary conditions, limiting the transversal dimensions of the modal fields. In the following we will use notation for the case where the modal fields have finite transversal dimensions; the case where the modal fields extend infinitely in the transversal direction is described by substituting sums by integrals, and Kronecker deltas by Dirac delta functions.

At the surface $s$ we can write the transversal components of the electric and magnetic fields as follows:
\begin{subequations}\label{sfields}
\begin{align}
\vek E^{\text{t}} &= \sum_j \left( a_j\vek e_j^{\text{t}} + b_j\vek e_j^{\text{t}} \right), \\
\vek H^{\text{t}} &= \sum_j \left( a_j\vek h_j^{\text{t}} - b_j\vek h_j^{\text{t}} \right). 
\end{align}
\end{subequations}
Here $\{\vek e_j,\vek h_j\}$ denotes the modal fields of mode $j$, and the superscript $\text{t}$ stands for the transversal component. The sums run over all modes in all waveguides. Note the minus sign of the backward propagating magnetic field; this is due to the usual mode convention in the literature (see Appendix \ref{waveguidemodes}). In \eqref{sfields} the spatial dependence $\exp(\pm i\beta_j z_j)$, where $\beta_j$ is the propagation constant of mode $j$, and $z_j$ is the associated waveguide axis, has been absorbed into the modal amplitudes $a_j$ and $b_j$.

Since the system in $v$ is linear, we can express the output propagating modes using a scattering matrix $S_{jk}$:
\begin{equation}
b_j=\sum_k S_{jk}a_k.
\end{equation}

Defining
\begin{equation}\label{orthogonality}
\int_s \vek e_j \times \vek h_k \cdot\diff\vek S = C_{jk},
\end{equation}
the properties of the matrix $C_{jk}$ are found to be as follows: When $j$ and $k$ refer to modes in different waveguides,  orthogonality $C_{jk}=0$ is trivially fulfilled since the associated modal fields are assumed to be non-overlapping. For the nontrivial case with $j$ and $k$ in the same waveguide, it is proved in Appendix \ref{waveguidemodes} that $C_{jk}$ is symmetric, and that $C_{jk}=0$ for modes $j$ and $k$ with distinct propagation constants. Thus $C_{jk}$ is a block-diagonal, symmetric matrix. We here note a generalization compared to earlier treatments in the literature: Due to possible nonorthogonality of degenerate modes, we permit $C_{jk}$ to be nondiagonal.

We now consider two different excitations and their associated fields on $s$. The first has $a_j^1=1$ for a certain $j$, putting all other $a_j^1$'s to zero; and the other has $a_k^2=1$ for a certain $k$, putting all other $a_k^2$'s to zero. Here the superscripts 1 and 2 denote the two different excitations.
This means that
\begin{subequations}\label{EtHt1}
\begin{align}
\vek E^{1\text{t}} &= \vek e_j^{\text{t}} + \sum_l S_{lj}\vek e_l^{\text{t}}, \\
\vek H^{1\text{t}} &= \vek h_j^{\text{t}} - \sum_l S_{lj}\vek h_l^{\text{t}}, 
\end{align}
\end{subequations}
and
\begin{subequations}\label{EtHt2}
\begin{align}
\vek E^{2\text{t}} &= \vek e_k^{\text{t}} + \sum_l S_{lk}\vek e_l^{\text{t}}, \\
\vek H^{2\text{t}} &= \vek h_k^{\text{t}} - \sum_l S_{lk}\vek h_l^{\text{t}}. 
\end{align}
\end{subequations}
Substituting \eqref{EtHt1} and \eqref{EtHt2} in the Lorentz reciprocity theorem \eqref{lorentzrec}, and using \eqref{orthogonality} yields
\begin{align}
& C_{jk}-\sum_{l,m}C_{lm}S_{lj}S_{mk}+\sum_l \left(C_{lk}S_{lj}-C_{jl}S_{lk}\right) \\
& = C_{kj}-\sum_{l,m}C_{ml}S_{lj}S_{mk}+\sum_l \left(C_{lj}S_{lk}-C_{kl}S_{lj}\right). \nonumber
\end{align}
Using the symmetry of $C_{jk}$ we conclude that
\begin{equation}\label{CSsymsum}
 \sum_l C_{jl}S_{lk} = \sum_l C_{kl}S_{lj}, 
\end{equation}
for all $j$ and $k$, i.e., the matrix $CS$ is symmetric: 
\begin{equation}\label{CSsym}
 CS = (CS)^\text{T}.
\end{equation}

In some cases of interest, we may have orthogonalized part of the mode set, while we do not care about the remaining part. This may for instance be the case when considering certain bound modes in dielectric waveguides, or when some of the input/output waveguides are irrelevant to the problem under investigation. For $j$ and/or $k$ in this restricted, orthogonal set, we can write (Appendix \ref{waveguidemodes}):
\begin{equation}\label{orthogonalitydiag}
\int_s \vek e_j \times \vek h_k \cdot\diff\vek S = c_{j}\delta_{jk},
\end{equation}
where the symbol $\delta_{jk}$ stands for the Kronecker delta, and $c_j$ is a normalization constant. In other words, $C_{jk}=c_j\delta_{jk}$ for $j$ and/or $k$ in the orthogonal set. By substitution into \eqref{CSsymsum} we get the important result: When mode $j$ and mode $k$ are orthogonal to each other and to all other modes,
\begin{equation}\label{ssymdiag}
c_{j}S_{jk} = c_kS_{kj}.
\end{equation}
We note that \eqref{ssymdiag} is valid even though there are nonorthogonal modes in the remaining part of the modal spectrum.

One common normalization choice is to let 
\begin{equation}\label{normalization}
c_j=\begin{cases}
     1, & \text{for propagating modes,} \\
     \pm i & \text{for evanescent modes,}
    \end{cases}
\end{equation}
while modes with complex or zero propagation constant may have general complex $c_j$. Then, when modes $j$ and $k$ are orthogonal to each other and to all other modes,
\begin{equation}
S_{jk}=S_{kj}
\end{equation}
for propagating modes, while 
\begin{equation}
S_{jk}=\pm iS_{kj}
\end{equation}
if $j$ is propagating and mode $k$ is evanescent.

We now consider some common special cases of the general reciprocity relation \eqref{CSsym}.

\section{Examples}

\subsection{Orthogonal waveguide modes on the left and free space on the right}
For the special case where one of the involved waveguides vanishes, i.e., we have free space on the right-hand side of the volume $v$ in Fig. \ref{fig:reciprocity}(b), we can take the modal fields on the right-hand side to be complex exponentials, or plane waves:
\begin{subequations}\label{complexexp}
\begin{align}
\vek e_j &= \vek{\hat e}_j\exp(ik_xx+ik_yy), \\
\vek h_j &= \vek{\hat h}_j\exp(ik_xx+ik_yy).
\end{align}
\end{subequations}
Here the constant vectors $\vek{\hat e}_j$ and $\vek{\hat h}_j$ express the polarization. Using Maxwell's equations we find the connection between these constant vectors, and also the dispersion relation
\begin{equation}\label{kz}
\beta^2=k_z^2=\epsilon\mu\frac{\omega^2}{c^2}-k_x^2-k_y^2. 
\end{equation}
Here $\epsilon$ and $\mu$ denote the relative permittivity and permeability, respectively, and $\omega$ is the angular frequency. The modes are labeled by $k_x$, $k_y$, and the polarization pol = TE or TM, so when $j$ and $k$ refer to modes in free space, they can be seen as a collection of indices, $j\to(k_x, k_y,\text{pol})$ and $k\to(k_x', k_y',\text{pol}')$. For convenience we let $-j\to(-k_x, -k_y,\text{pol})$ and $-k\to(-k_x', -k_y',\text{pol}')$.

We take $\vek{\hat h}_j^\text{t}$ to be real. Although not assumed here, for real $\epsilon$ and $\mu$ this means that $\vek{\hat e}_j^\text{t}$ is real for propagating modes, and imaginary for evanescent modes. Using an appropriate mode normalization, we find the orthogonality relation
\begin{align}\label{orthogonalityexp}
& \int \vek e_{(k_x,k_y,\text{pol})} \times \vek h_{(-k_x',-k_y',\text{pol}')} \cdot\diff\vek S \nonumber\\ 
& = c(k_z,\text{pol}) \delta(k_x-k_x')\delta(k_y-k_y')\delta_{\text{pol}',\text{pol}},
\end{align}
by direct calculation. Here the integration extends over the entire $xy$-plane, and
\begin{equation}
c(k_z,\text{pol})=\begin{cases}
        \frac{\mu}{|\mu|}\frac{|k_z|}{k_z}, & \text{ for pol}=\text{TE}, \\
        \frac{|\epsilon|}{\epsilon}\frac{k_z}{|k_z|}, & \text{ for pol}=\text{TM},\\
        0, & \text{ for $k_z=0$.}
       \end{cases}
\end{equation}

Taking the other waveguides into account, and assuming orthogonal modes in them, we obtain \eqref{orthogonality}, with
\begin{equation}
C_{jk}=\begin{cases}
     c(k_z,\text{pol})\delta_{j(-k)}, & \text{$j$ and $k$ refer to free space modes,} \\
     c_j\delta_{jk}, & \text{$j$ and $k$ refer to waveguide modes,} \\
     0, & \text{otherwise.}
    \end{cases} 
\end{equation}
The short hand notation $\delta_{j(-k)}$ for $j=(k_x,k_y,\text{pol})$ and $k=(k_x',k_y',\text{pol}')$ means
\begin{equation}
\delta_{j(-k)} =  \delta(k_x+k_x')\delta(k_y+k_y')\delta_{\text{pol},\text{pol}'}.
\end{equation}
Due to the nonorthogonality of the degenerate modes $(k_x,k_y,\text{pol})$ and $(-k_x,-k_y,\text{pol})$, $C$ is not diagonal in this case, so we invoke the general version \eqref{CSsymsum} of the symmetry relation. This gives (here we note that the sum is replaced by an integral for the continuous spectrum of free space modes):
\begin{subequations}
\label{recfreespacemixed}
\begin{align}
& c_jS_{jk}=c_kS_{kj}, \\
& c_jS_{j,(k_x,k_y,\text{pol})}=c(k_z,\text{pol}) S_{(-k_x,-k_y,\text{pol}),j}, \label{transmwgfs}\\
& c(k_z,\text{pol}) S_{(k_x,k_y,\text{pol}),(k_x',k_y',\text{pol}')} \nonumber\\
&=c(k_z',\text{pol}') S_{(-k_x',-k_y',\text{pol}'),(-k_x,-k_y,\text{pol})}, \label{recfreespace}
\end{align}   
\end{subequations}
In \eqref{recfreespacemixed} $j$ and $k$ denote waveguide modes, while $(k_x, k_y, \text{pol})$ and $(k_x', k_y', \text{pol}')$ denote free-space modes on the right-hand side of the scatterer. Moreover
\begin{equation}\label{kzp}
k_z'^2=\epsilon\mu \omega^2-k_x'^2-k_y'^2.                                                                                                                                                                                                                                                                                                                                                                                                                                                       \end{equation}

Instead of using the complex exponentials \eqref{complexexp} as the modes, we can form linear combinations of modes with opposite signs of $k_x$, and opposite signs of $k_y$. This gives modal fields with components of the form $\cos(k_xx)\cos(k_yy)$, $\cos(k_xx)\sin(k_yy)$, $\sin(k_xx)\cos(k_yy)$, or $\sin(k_xx)\sin(k_yy)$. It is straightforward to verify that the resulting modes are orthogonal, i.e., $C$ now becomes diagonal. Thus in this case we obtain
\begin{equation}
\label{recfreespacemixedsincos}
c_jS_{jk}=c_kS_{kj},
\end{equation}
rather than \eqref{recfreespacemixed}. Here $j$ and $k$ may denote waveguide or free space modes. For free space modes $c_j=c(k_z,\text{pol})$ and $c_k=c(k_z',\text{pol}')$.

\subsection{Free space on both sides}
When there is free space on both sides of the scatterer, the analysis is similar to that in the previous subsection. We find that \eqref{recfreespace} remains valid for the reflection coefficients on each side when the modal fields are taken to be complex exponentials. If there are different homogeneous media on each side, the appropriate electromagnetic parameters must be used in \eqref{kz} and \eqref{kzp}. For the transmission coefficients we obtain
\begin{align}
& c(k_z,\text{pol}) S_{(\text{left},k_x,k_y,\text{pol}),(\text{right},k_x',k_y',\text{pol}')} \nonumber\\
&= c(k_z',\text{pol}') S_{(\text{right},-k_x',-k_y',\text{pol}'),(\text{left},-k_x,-k_y,\text{pol})}, \label{recfreespacebothsides} 
\end{align}
where ``left'' and ``right'' specify whether the particular mode is located on the left-hand side or right-hand side of the scatterer. This special case was treated previously in Refs. \cite{carminati1998,carminati2000}.

If the sin/cos type modes are used instead, \eqref{recfreespacemixedsincos} remains valid.

\subsection{Discontinuity between two waveguides}
\begin{figure}
\includegraphics[width=8cm]{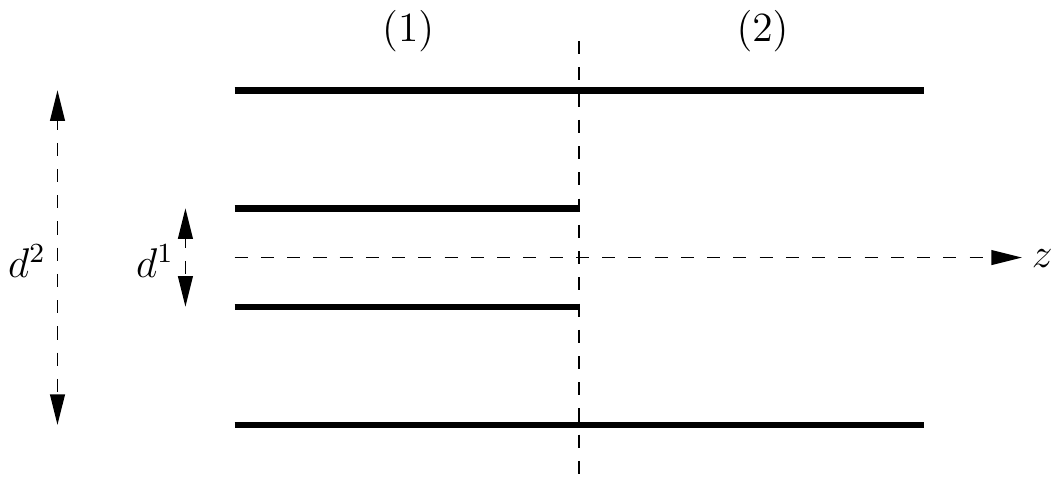}
\caption{The discontinuity between parallel plate, metallic waveguides. On the left-hand side (1) there is a waveguide of thickness $d^1$ inserted into another waveguide of thickness $d^2$. On the right-hand side (2), only the waveguide of thickness $d^2$ is present.}
\label{fig:metallic}
\end{figure}

Consider the discontinuity between two waveguides, see, for example, Fig. \ref{fig:metallic}. The goal is to calculate the reflection coefficients and transmission coefficients from either side, demonstrating that they obey the reciprocity relation \eqref{ssymdiag}. For completeness and notational consistency, we include a derivation based on Refs. \cite{svendsen2011,svendsen2012}.

First we let the two waveguides be arbitrary. However, for simplicity we use sum notation for the superposition of modes, indicating that any continuous part of the mode spectrum is discretized using artificial boundary conditions. Furthermore we assume that the mode sets are orthogonal, i.e. they satisfy
\begin{subequations}\label{orthogonalitydiagwg}
\begin{align}
\int_A \vek e_j^1 \times \vek h_k^1 \cdot\diff\vek S = c_{j}^1\delta_{jk}, \\
\int_A \vek e_j^2 \times \vek h_k^2 \cdot\diff\vek S = c_{j}^2\delta_{jk},
\end{align}
\end{subequations}
where $A$ is the boundary between the two waveguides, and $c_j^{1,2}$ denotes the normalization constant of mode $j$ in waveguide $1,2$ (see \eqref{eq:ortodiag}). Define the overlaps
\begin{subequations}\label{overlap}
\begin{align}
\Psi_{jk}^1 = \int_{A} \vek e^1_j\times\vek h^2_k\cdot\diff\vek S, \\
\Psi_{jk}^2 = \int_{A} \vek e^2_k\times\vek h^1_j\cdot\diff\vek S.
\end{align}
\end{subequations}

We consider the situation where mode $j$ is incident from waveguide 1 to waveguide 2, and impose continuity of the transversal electric and magnetic fields. This amounts to solving the equations
\begin{subequations}\label{matcheh}
\begin{align}
\vek e_j^{1\text t} + \sum_k r_{kj}^1\vek e_k^{1\text t} &= \sum_k t_{kj}^1 \vek e_k^{2\text t}, \\
\vek h_{j}^{1\text t} - \sum_k r_{kj}^1\vek h_{k}^{1\text t} &= \sum_k t_{kj}^1 \vek h_{k}^{2\text t}, 
\end{align}
\end{subequations}
where $r_{kj}^1$ is the reflection coefficient from mode $j$ to mode $k$, as seen from waveguide 1, and $t_{kj}^1$ is the transmission coefficient from mode $j$ in waveguide 1 to mode $k$ in waveguide 2. Using \eqref{orthogonalitydiagwg} and \eqref{overlap} we can rewrite \eqref{matcheh}:
\begin{subequations}\label{PsiPhir1}
\begin{align}
c_j^1\delta_{jm} + c_m^1 r_{mj}^1 &= \sum_k t_{kj}^1\Psi^2_{mk}, \\
c_j^1\delta_{jm} - c_m^1 r_{mj}^1 &= \sum_k t_{kj}^1\Psi^1_{mk}. 
\end{align}
\end{subequations}
Summing these equations we find
\begin{equation}\label{transmeq1}
\sum_k(\Psi_{mk}^1+\Psi^{2}_{mk})t^1_{kj}/c_j^1=2\delta_{jm}. 
\end{equation}
This matrix equation can be inverted, yielding a matrix Fresnel equation expressing the transmission coefficients $t^1_{kj}$ from the modal fields.

Repeating the derivation above, we find the transmission coefficients from waveguide 2 to waveguide 1 by solving
\begin{equation}\label{transmeq2}
\sum_k(\Psi_{km}^1+\Psi^{2}_{km})t^2_{kj}/c_j^2=2\delta_{jm}. 
\end{equation}
Comparing \eqref{transmeq1} and \eqref{transmeq2} as matrix equations, we realize that 
\begin{equation}
c_j^1 t^2_{jk}=c_k^2 t^1_{kj}, 
\end{equation}
in agreement with the reciprocity relation \eqref{ssymdiag}.

The reflection matrix can be found by eliminating the transmission coefficients from \eqref{PsiPhir1}, or alternatively, by formulating \eqref{matcheh} as follows:
\begin{subequations}\label{PsiPhit1}
\begin{align}
\Psi^1_{jm} + \sum_k r_{kj}^1\Psi^1_{km} &= c_m^2t_{mj}^1, \\
\Psi^2_{jm} - \sum_k r_{kj}^1\Psi^2_{km} &= c_m^2t_{mj}^1. 
\end{align}
\end{subequations}
In matrix notation, we define $R^1=[r^1_{kj}]$ and $\Psi^{1,2}=[\Psi^{1,2}_{jm}]$. Eqs. \eqref{PsiPhit1} can then be combined to the matrix Fresnel equation
\begin{equation}\label{matrixR1}
R^{1\text T}=(\Psi^2-\Psi^1)(\Psi^1+\Psi^2)^{-1}. 
\end{equation}
Similarly we find the reflection matrix as seen from waveguide 2:
\begin{equation}
R^{2}=(\Psi^1+\Psi^2)^{-1}(\Psi^1-\Psi^2). 
\end{equation}

To demonstrate that these expressions are consistent with reciprocity \eqref{ssymdiag}, we specialize to two planar, nonmagnetic waveguides, and an incident TE mode. Both waveguides are oriented such that the structure is uniform in the $y$-direction. Since there will be no coupling between TE and TM modes, we can disregard TM modes from the analysis. By straightforward calculation we find
\begin{subequations}\label{Psi12}
\begin{align}
\Psi^1_{jk}=\beta^2_k\Psi_{jk}, \\
\Psi^2_{jk}=\beta^1_j\Psi_{jk},
\end{align}
\end{subequations}
where $\beta^1_j$ ($\beta^2_k$) is the propagation constant of mode $j$ ($k$) in waveguide 1 (2), and
\begin{equation}\label{Psi}
\Psi_{jk}=\frac{1}{\omega\mu_0}\int_\text{cross section} \vek e_j^{1\text t}\cdot \vek e_k^{2\text t}\diff x. 
\end{equation}
From \eqref{Psi} it is apparent that even and odd modes do not couple to each other, so we limit the discussion to even modes.

\begin{figure}
\includegraphics[width=8cm]{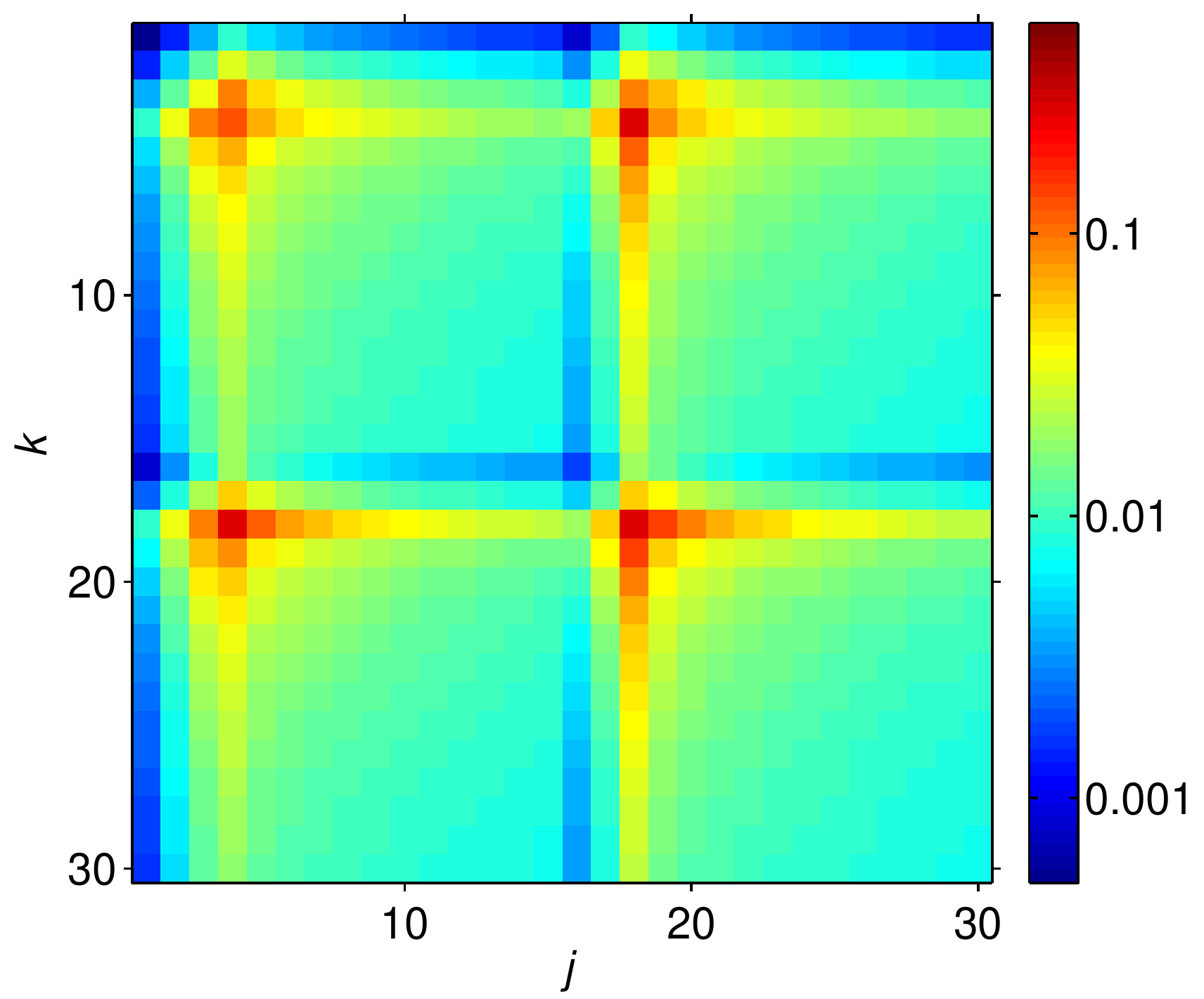}
\caption{The matrix $|c_k r^1_{kj}|$. The numbers used are $d^1=d$, $d^2=2d$, and $\omega d/c=20$. In the plot, mode $1, 2, \ldots, 15$ correspond to even TE modes in the central waveguide on the left-hand side, while $16, \ldots, 30$ correspond to even TE modes in the upper and lower waveguides. Modes $1, 2, 3, 16, 17, $ and $18$ are propagating, while the remaining modes are evanescent. The plot was generated using 100 modes on each side in the calculations, which leads to an asymmetry of $10^{-3}$ for the matrix. This error is reduced to $10^{-5}$ by using 1000 modes.}
\label{fig:reflectionmetallic}
\end{figure}

First we consider the situation where the waveguides are parallel plate, metallic waveguides, with thicknesses $d^1$ and $d^2$, respectively, and common axis $\vekh z$, see Fig. \ref{fig:metallic}. The waveguides are filled with vacuum. The modes are easily found analytically, yielding analytic expressions for $\beta^1_j$, $\beta^2_k$, and $\Psi_{jk}$. Using \eqref{matrixR1} and \eqref{Psi12}, the resulting matrix $c_k^1 r^1_{kj}$ is calculated and shown in Fig. \ref{fig:reflectionmetallic}.
We observe that the matrix is symmetric, as predicted by reciprocity: Eq. \eqref{ssymdiag} can in this case be written
\begin{equation}
c_j^1 r^1_{jk}=c_k^1 r^1_{kj}.
\end{equation}
Using the normalization \eqref{normalization}, $|c_k|=1$ for all $k$, which means that $|r_{kj}|$ becomes symmetric.

\begin{figure}
\includegraphics[width=8cm]{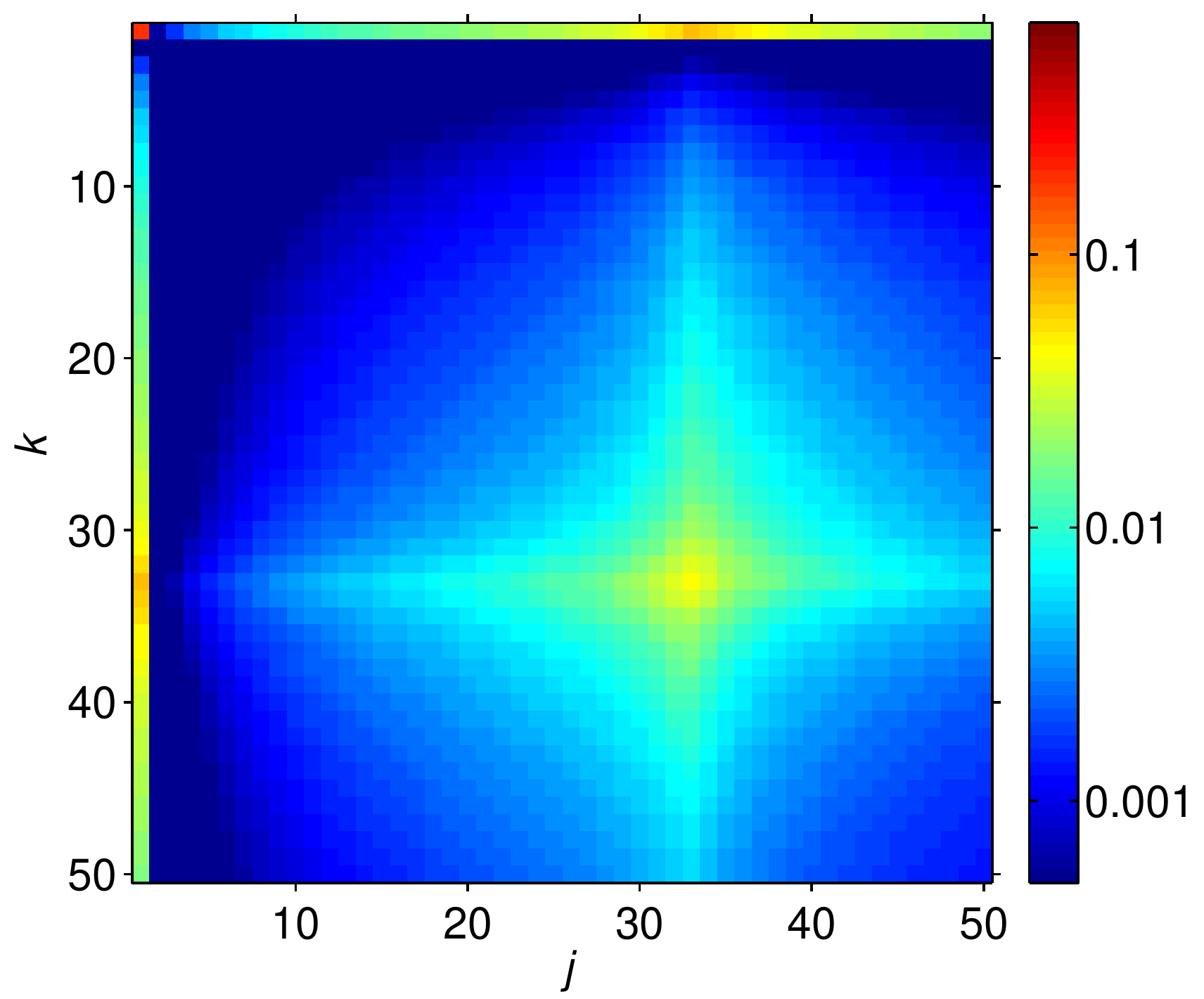}
\caption{The matrix $|c_k r^1_{kj}|$. The refractive index of the dielectric waveguide core is set to 1.5, and $\omega d/c=1$. The modes are computed numerically using periodic boundary conditions with period $L=100d$. To reduce the effect of the artificial boundary conditions, we introduce a small loss everywhere, such that $\epsilon\to\epsilon+i0.05$. The coefficients for the 50 lowest order TE waveguide modes are displayed in the figure. Only the first waveguide mode is bound, modes $2, \ldots, 31$ are unbound and propagating (i.e., with small imaginary parts of the propagation constants), and the remaining modes are unbound and evanescent (i.e., with large imaginary parts of the propagation constants). The plot was generated using 100 modes on the waveguide side. On the free space side, for simplicity we used complex exponentials as the modal fields. Since these fields are neither even nor odd, we included twice as many modes there, corresponding to positive and negative transverse wavenumbers. When the matrix Fresnel equation (\ref{matrixR1}) is overdetermined, the matrix inverse should be replaced by the Moore--Penrose pseudoinverse. Using 100 waveguide modes leads to a matrix asymmetry of $10^{-3}$. Again, the error is reduced to $10^{-5}$ when using 1000 waveguide modes.}
\label{fig:reflectiondielectric}
\end{figure}

\begin{figure}
\includegraphics[width=8cm]{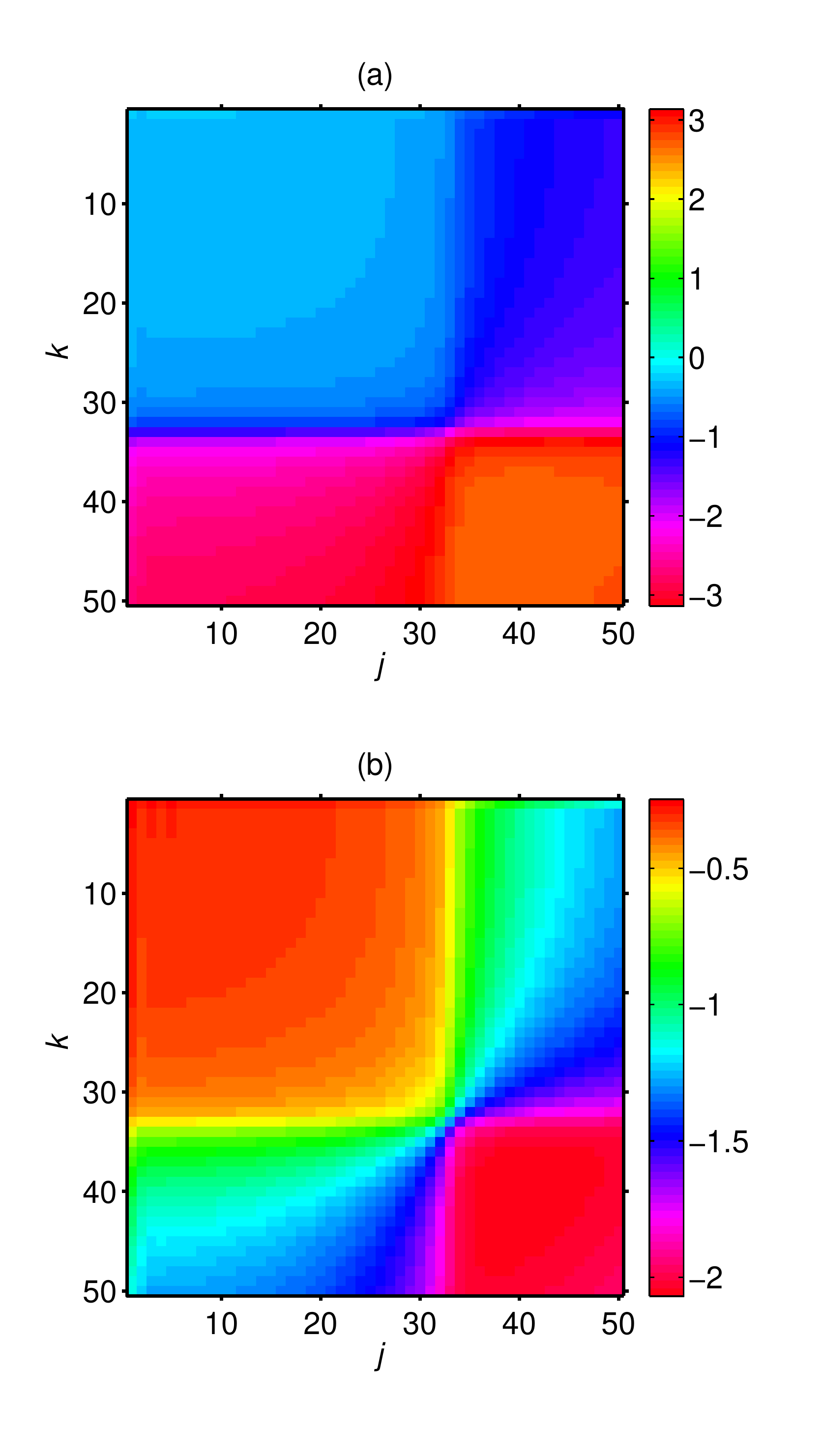}
\caption{(a) Phase of $r^1_{kj}$, and (b) phase of $c_kr^1_{kj}$, for the dielectric waveguide in Fig.~\ref{fig:reflectiondielectric}. The phase plots clearly show that multiplication by the $c_{k}$ coefficients is necessary to obtain a symmetric matrix.}
\label{fig:dielectric_phase}
\end{figure}

Second we consider the end facet of a thin, dielectric, planar waveguide of thickness $d$. Using \eqref{matrixR1}, \eqref{Psi12}, and \eqref{Psi} we compute the reflection matrix $r^1_{kj}$. The resulting matrix $|c_kr^1_{kj}|$ is depicted in Fig. \ref{fig:reflectiondielectric}, and is clearly symmetric. In Fig.~\ref{fig:dielectric_phase}(a), we also show the phase of $r^1_{kj}$, which demonstrates the asymmetry of the scattering matrix due to presence of evanescent modes. For comparison, Fig.~\ref{fig:dielectric_phase}(b) shows the phase of the symmetric matrix $c_kr^1_{kj}$.

\section{Conclusion}
We have considered the implications of the Lorentz reciprocity theorem for the scattering matrix. The modes to and from the scatterer are described as general waveguide modes, including propagating, evanescent, and complex modes. The modes may also be degenerate and nonorthogonal. Despite this generality, a simple reciprocity relation \eqref{CSsym} can be derived, stating that the matrix product of the orthogonality matrix $C$ (with elements defined by \eqref{orthogonality}) and the scattering matrix $S$, is symmetric. This symmetry relation is expected to be useful in photonics and microwave engineering, when designing structures involving complex waveguides.

\appendix
\section{Lorentz' reciprocity theorem}
Consider a volume $v$ with a linear, time-invariant, and spatially nondispersive medium. The conductivity, if any, is included in the (possibly complex) permittivity tensor $\epsilon$. We assume that there are no current sources in $v$, and that the tensors $\epsilon$ and $\mu$ are symmetric. Let \{$\vek E^1$,$\vek H^1$\} and \{$\vek E^2$,$\vek H^2$\} be the electromagnetic fields in $v$, resulting from excitations 1 and 2, respectively. From Maxwell's equations we have (assuming time dependence $\exp(-i\omega t)$):
\begin{align}
\nabla\cdot(\vek E^1\times\vek H^2)&=-\vek E^1\cdot\curl H^2+\vek H^2\cdot\curl E^1 \nonumber\\ 
&=i\omega\vek E^1\cdot(\epsilon\vek E^2)+i\omega\vek H^2\cdot(\mu\vek H^1)
\end{align}
and
\begin{equation}
\nabla\cdot(\vek E^2\times\vek H^1)=i\omega\vek E^2\cdot(\epsilon\vek E^1)+i\omega\vek H^1\cdot(\mu\vek H^2),
\end{equation}
Using the symmetry of $\epsilon$, we have
\begin{equation}
\vek E^1\cdot (\epsilon\vek E^2)=\sum_{k,l=1}^3 E^1_k\epsilon_{kl}E^2_l=\sum_{k,l=1}^3 E^2_l\epsilon_{lk}E^1_k=\vek E^2\cdot (\epsilon\vek E^1).
\end{equation}
Similarly,
\begin{equation}
\vek H^1\cdot (\mu\vek H^2)=\vek H^2\cdot (\mu\vek H^1).
\end{equation}
We therefore obtain the usual Lorentz reciprocity theorem:
\begin{equation}
 \nabla\cdot(\vek E^1\times\vek H^2)=\nabla\cdot(\vek E^2\times\vek H^1),
\end{equation}
or, in integral form:
\begin{equation}
 \oint_s (\vek E^1\times\vek H^2)\cdot\diff\vek S=\oint_s (\vek E^2\times\vek H^1)\cdot\diff\vek S,
\end{equation}
where the closed surface $s$ is the boundary of $v$.

\section{Waveguide modes}\label{waveguidemodes}
For structures that are homogeneous in the $z$-direction, the solutions to Maxwell's equations are expressible as superpositions of waveguide modes, that is, solutions in the form
\begin{subequations}
\begin{align}
\vek E &=a(\vek e^{\text t}+e^z\vekh z)\exp(i\beta z), \\
\vek H &=a(\vek h^{\text t}+h^z\vekh z)\exp(i\beta z).
\end{align}
\end{subequations}
Here the superscripts t and $z$ stand for the vector components perpendicular and parallel to the $z$-axis, respectively, and $a$ is a constant. Assuming a (possibly lossy) medium with permittivity $\epsilon$ and permeability $\mu$, and substituting into the source-free Maxwell equations, one obtains
\begin{subequations}\label{maxwellmodes}
\begin{align}
\vek e^{\text t} &=-\frac{1}{\omega\epsilon} \vekh z\times \left(\beta\vek h^{\text t}+i\nabla^\text{t} h^z \right), \label{eq:et}\\
\vek h^{\text t} &=\frac{1}{\omega\mu} \vekh z\times \left(\beta\vek e^{\text t}+i\nabla^\text{t} e^z \right), \label{eq:ht}\\
e^{z} &=\frac{i}{\omega\epsilon} \vekh z\cdot \nabla^\text{t}\times \vek h^\text{t}, \label{eq:ez}\\
h^{z} &=-\frac{i}{\omega\mu} \vekh z\cdot \nabla^\text{t}\times \vek e^\text{t}.\label{eq:hz}
\end{align}
\end{subequations}
It is therefore consistent to choose ``backward propagating modes'' as related to ``forward propagating modes'' as follows: Under the transformation $\beta\to -\beta$, we require $\vek e^{\text t} \to \vek e^{\text t}$, $e^z \to -e^z$, $h^z \to h^z$, and $\vek h^{\text t} \to -\vek h^{\text t}$.

By inserting Eq. (\ref{eq:hz}) in Eq. (\ref{eq:et}), and Eq. (\ref{eq:ez}) in Eq. (\ref{eq:ht}), and rearranging the terms, Maxwell's equations take the form of a generalized eigenproblem \cite{bresler1958,johnson2002}
\begin{equation}\label{eq:eigeneq}
\hat{A}|\Psi\rangle=\beta\hat{B}|\Psi\rangle,
\end{equation}
where
\begin{equation}
\hat{A}=\matrise{\omega\epsilon-\frac{1}{\omega}\nabla^\text{t}\times\frac{1}{\mu}\nabla^\text{t}\times & 0 \\ 0 & \omega\mu-\frac{1}{\omega}\nabla^\text{t}\times\frac{1}{\epsilon}\nabla^\text{t}\times},
\end{equation}
\begin{equation}
\hat{B}=\matrise{0 & -\vekh z\times \\ \vekh z\times & 0},
\end{equation}
and
\begin{eqnarray}
|\Psi\rangle=\matrise{\vek e^\text{t} \\ \vek h^\text{t}}.
\end{eqnarray}
The inner product between two vectors $|\Psi_j\rangle$ and $|\Psi_k\rangle$ is defined as
\begin{equation}\label{innerproddef}
\langle\Psi_j|\Psi_k\rangle=\frac{1}{2}\int_A \left[(\vek e^{\text{t}}_j)^*\cdot\vek e^\text{t}_k+(\vek h^{\text{t}}_j)^*\cdot\vek h^\text{t}_k\right]\diff S,
\end{equation}
where the integration area $A$ either is the entire $xy$-plane or the computational area associated with, for example, periodic or metallic boundary conditions. It is straightforward to prove that the operator $\hat{B}$ is hermitian and real, and therefore symmetric. Also $\hat B^{-1}=\hat B$. For general lossy media, $\hat{A}$ is not hermitian; however we can prove that it is symmetric (i.e., $\hat A^{*\dagger}=\hat A$) by verifying the equation $\langle\Psi_j|\hat A|\Psi_k\rangle=(\hat A^* |\Psi_j\rangle)^\dagger|\Psi_k\rangle$. Indeed, $\div(\vek u\times\vek v)=(\curl u)\cdot\vek v-\vek u\cdot(\curl v)$, and therefore
\begin{equation}\label{condHerm}
\int_A (\curl u)\cdot\vek v \diff S = \int_A (\curl v)\cdot\vek u \diff S
\end{equation}
whenever the line integral $\oint_{\partial A}(\vek u\times\vek v)\cdot\vekh n\diff l=0$. Here $\vekh n$ is the outward-pointing unit normal vector to the boundary $\partial A$ of $A$. This condition is satisfied for metallic or periodic boundary conditions, and can be justified in general by enclosing the waveguide by a conducting cylinder whose radius approaches infinity.

Transposing \eqref{eq:eigeneq}, and using the symmetry of $\hat A$ and $\hat B$, we obtain
\begin{equation}\label{eq:eigeneq2}
\langle\Psi^*|\hat{A}=\beta\langle\Psi^*|\hat{B}.
\end{equation}
By considering (\ref{eq:eigeneq}) for a waveguide mode $|\Psi_k\rangle$, taking the inner product with $|\Psi_j^*\rangle$, we obtain
\begin{equation}\label{eq:ort2}
(\beta_j-\beta_k)\langle\Psi_j^*|\hat{B}|\Psi_k\rangle=0.
\end{equation}
Thus if $\beta_j\neq\beta_k$ we have
\begin{equation}\label{eq:ort3}
\langle\Psi_j^*|\hat{B}|\Psi_k\rangle=0.
\end{equation}
With the help of \eqref{innerproddef}, Eq. (\ref{eq:ort3}) can be written
\begin{equation}\label{eq:orto1}
\frac{1}{2}\int_A \left(\mathbf{e}_j\times\mathbf{h}_k+\mathbf{e}_k\times\mathbf{h}_j\right)\cdot\mathbf{\hat{z}}\text{d}S=0
\end{equation}
for $\beta_j\neq\beta_k$. Using the transformation properties of $\mathbf{e}_k$ and $\mathbf{h}_k$ as $\beta_k\rightarrow -\beta_k$, we obtain
\begin{equation}\label{eq:orto2}
\frac{1}{2}\int_A \left(-\mathbf{e}_j\times\mathbf{h}_k+\mathbf{e}_k\times\mathbf{h}_j\right)\cdot\mathbf{\hat{z}}\text{d}S=0
\end{equation}
for $\beta_j\neq-\beta_k$. Thus, for $\beta_j\neq\pm\beta_k$ we obtain
\begin{equation}\label{eq:ortofin}
\int_A \mathbf{e}_j\times\mathbf{h}_k\cdot\mathbf{\hat{z}}\text{d}S=0,
\end{equation}
by subtracting (\ref{eq:orto2}) from (\ref{eq:orto1}). Note that the orthogonality \eqref{eq:ortofin} is valid for all nondegenerate waveguide modes, including complex modes. For degenerate modes, however, we are not guaranteed that \eqref{eq:ortofin} is fulfilled, even though the mode set may in some cases be orthogonalized, for example using the Gram--Schmidt procedure\footnote{The Gram--Schmidt procedure requires the modes to be normalizable, i.e., 
$\langle\Psi^*|\hat B|\Psi\rangle\neq 0$ for $|\Psi\rangle$ in the degeneracy space. This is not always possible, as exemplified by the modes $\vek e=\exp(ik_xx)\vekh y$ in free space: In the sense \eqref{eq:ort3} these modes are orthogonal with themselves, and therefore not normalizable. On the other hand they are not orthogonal to their counterparts $\vek e=\exp(-ik_xx)\vekh y$. Even though the Gram--Schmidt procedure is not directly applicable, the mode set can be orthogonalized by the superpositions $2\cos(k_xx)=\exp(ik_xx)+\exp(-ik_xx)$ and $2i\sin(k_xx)=\exp(ik_xx)-\exp(-ik_xx)$. Note, however, that such orthogonalization is not always desirable; in the example complex exponentials are often more convenient to work with than sin/cos type modes.}.
In general we therefore write
\begin{equation}\label{eq:ortodeg}
\int_A \mathbf{e}_j\times\mathbf{h}_k\cdot\mathbf{\hat{z}}\text{d}S=C_{jk},
\end{equation}
where $C_{jk}$ is a block diagonal matrix such that $C_{jk}=0$ for nondegenerate modes ($\beta_j\neq\pm\beta_k$). From \eqref{eq:orto2} we find that $C_{jk}=C_{kj}$ provided $\beta_j\neq -\beta_k$. The latter condition can always be assumed fulfilled by restricting the mode set to forward propagating modes, i.e., by making a mode set consisting of the half of modes satisfying $\beta_j\neq -\beta_k$. For gainless waveguides this amounts to picking the modes with power flow in $+z$-direction (propagating modes), or $\Im\beta>0$ (evanescent and complex modes). For modes with real $\beta$ and zero power flow in the $z$-direction, the forward propagating subset can be found by considering a lossy waveguide in the lossless limit, or by invoking causality.

Provided the degenerate modes can be orthogonalized, we can write
\begin{equation}\label{eq:ortodiag}
\int_A \mathbf{e}_j\times\mathbf{h}_k\cdot\mathbf{\hat{z}}\text{d}S=c_{j}\delta_{jk}.
\end{equation}
In principle we can take $c_j=1$ for all $j$ by a suitable normalization. It is however conventional to let
\begin{equation}\label{normalizationcj}
c_j=\begin{cases}
     1, & \text{for propagating modes,} \\
     \pm i, & \text{for evanescent modes,}
    \end{cases}
\end{equation}
while $c_j$ is generally complex for complex modes, and zero for modes with $\beta_j=0$. This is done such that, if possible, $\vek h_k^\text t$ can be taken to be real. Then \eqref{eq:ortodiag} implies power orthogonality for these modes:
\begin{equation}\label{eq:ortodiagcc}
\int_A \mathbf{e}_j\times\mathbf{h}_k^*\cdot\mathbf{\hat{z}}\text{d}S=c_{j}\delta_{jk}.
\end{equation}

\def\cprime{$'$}
%

\end{document}